\documentclass[%
 reprint,
 amsmath,amssymb,
 aps,
]{revtex4-1}

\usepackage{graphicx}
\usepackage{dcolumn}
\usepackage{bm}

\begin{document}

\preprint{APS/123-QED}

\title{A gauge-invariant and current-continuous microscopic ac quantum transport theory}

\author{JianQiao Zhang$^1$, ZhenYu Yin$^1$, Xiao Zheng$^2$, ChiYung Yam$^1$ and GuanHua Chen$^{1*}$}
\affiliation{
$^1$Department of Chemistry, the University of Hong Kong, Hong Kong, China\\
$^2$Hefei National Laboratory for Physical Sciences at the Microscale, University of Science and Technology of China, Hefei, Anhui 230026, China}

\begin{abstract}
There had been consensus on what the accurate ac quantum transport theory was until some recent works challenged the conventional wisdom. Basing on the non-equilibrium Green's function formalism for time-dependent quantum transport, we derive an expression for the dynamic admittance that satisfies gauge invariance and current continuity, and clarify the key concept in the field. The validity of our now formalism is verified by first-principles calculation of the transient current through a carbon-nanotube-based device under the time-dependent bias voltage. Moreover, the previously well-accepted expression for dynamic admittance is recovered only when the device is a perfect conductor at a specific potential.
\begin{description}
\item[PACS numbers]
72.10.Bg, 73.23.Ad, 73.63.-b, 73.63.Fg
\end{description}
\end{abstract}

\pacs{Valid PACS appear here}
\maketitle

Understanding quantum transport is important for the development of nanoelectronics. The Laudaure-B\"{u}ttiker formula has been widely used to calculate steady state current with great success.\cite{Xue2002,Datta2004,Taylor2001,Lang1995,Lang2000,Ke2004} The resulting steady state currents through the leads in contact with the electronic device satisfy gauge invariance and current conservation. Here, gauge invariance condition means that all physical observables should remain unchanged when the applied bias voltages are shifted by a constant. Current conservation implies that the steady state currents satisfy the Kirchhoff's circuit law, i.e., the currents through all leads sum up to zero. In contrast, the situation is somewhat different for ac quantum transport. The gauge invariance should still hold. Meanwhile, the particle currents through all leads are no longer conserved, due to the possible transient charge accumulation or depletion at the device. In such cases, the current conservation (or Kirchhoff's law) does not apply. Instead, the currents satisfy the charge continuity equation. Of course, the current conservation may be retrieved formally by introducing the displacement current, so that the total (particle plus displacement) currents through all leads sum up to zero. 

In 1993, B\"{u}ttiker and coworkers have derived an expression for the linear dynamic admittance of a small conductor by employing the scattering matrix theory.\cite{Buttiker1993} By taking into account of a uniform potential distribution inside the small conductor, gauge invariance was guaranteed. To ensure current conservation, the conductor was treated as an extra terminal, and the resulting additional dynamic admittance matrix counts for the contribution from the displacement current. However, it is unclear how to apply such a formulation to a generic device where the potential distribution is not uniform. In 1995, Anantram and Datta pioneered a microscopic ac transport theory based on nonequilibrium Green's function (NEGF) formalism and the wide-band limit (WBL) approximation.\cite{Anantram1995} However, the resulting formalism for the dynamic admittance does not satisfy gauge invariance and current conservation. To remedy such a situation, in 1999, a phenomenological NEGF method was proposed to ensure both gauge invariance and current conservation by Wang, Wang and Guo.\cite{Wang1999} The key idea was to partition the displacement current onto each electrode or terminal in a phenomenological manner. This was achieved by imposing the requirements of both gauge invariance and current conservation on top of the theory of Anantram and Datta.\cite{Anantram1995} It was a very nice idea and the resulting expression for the dynamic admittance has been widely used in the field ever since,\cite{Yamamoto2010} despite of the two facts: (1) it was built on a previous formalism that is not gauge-invariant;\cite{Anantram1995} and (2) no detailed justification or derivation based on the microscopic theory was given. 

We calculated the dynamic admittance of a carbon nanotube (CNT) based device, and compared it with the result of a well-established NEGF-based time-dependent density-functional theory (TDDFT-NEGF) method.\cite{Zheng2007,Zheng2010} The two sets of results are found significantly different from each other (see Fig.1b). Recently, Wang and coworkers have calculated the dynamic admittances of a benzene-dithiol and a chain of carbon atoms sandwiched between two aluminum electrodes.\cite{Zhuang2011} Both the phenomenological formalism of Reference\cite{Wang1999} and a microscopic theory\cite{Wei2009} were employed in their calculations. It was found that the discrepancy between the two could be as much as two orders of magnitude. The inconsistency in the calculation results thus presents a serious challenge for the existing ac quantum transport theory. 

Before we proceed to develop a correct microscopic theory for ac quantum transport, we explore what it might be. Current conservation for steady state current comes from the current continuity equation $\nabla\cdot\vec J +\frac{\partial}{\partial t}\rho=0$, where $\vec J(\vec r, t)$ is the current density function and $\rho(\vec r, t)$ the electron density function. For steady states, the time-partial derivative of $\rho(\vec r, t)$ is zero, so $\nabla\cdot\vec J=0$, which leads to current conservation. However, for ac transport, $\rho(\vec r, t)$ is time-dependent and its time-derivative is usually nonzero, and thus $\nabla\cdot\vec J\ne 0$, i.e. the current conservation may not hold. In other words, for ac or time-dependent current, the current conservation is not required. Certainly, if the displacement current is included, current conservation is satisfied for the total current, just as that in the conventional theory of electrodynamics.

As demonstrated by B\"{u}ttiker and coworkers, gauge invariance can be preserved by taking into account of the induced potential distribution of the electronic device properly.\cite{Buttiker1993} Due to the bias voltages on the electrodes, the electrons in the device region respond via redistribution. Moreover, electrons may leave from or enter into the device, which leads to further charge redistribution. The Hartree potential in the device then adjusts to the applied voltages and charge redistribution. As a result, the relative Hartree potential distribution is gauge-invariant, so is the resulting current, as required by physics. Therefore, the key to ensure gauge invariance is to self-consistently determine the Hartree potential.

Our objective is thus to develop an ac quantum transport theory that is gauge-invariant and current-continuous (rather than current-conserved). Starting from the Keldysh NEGF formula in time domain,\cite{Jauho1994}
\begin{equation}
\begin{split}
I_\alpha(t)&=\frac{e}{\hbar}\int_{-\infty}^{+\infty}{\rm dt_1}Tr[G^<(t, t_1)\Sigma_\alpha^a(t_1, t)\\
&+G^a(t, t_1)\Sigma_\alpha^<(t_1, t)+h. c.],
\end{split}
\end{equation}
we perform a double-time Fourier transform to convert it into frequency domain. The Green's function and self-energies are separated into equilibrium and small-signal terms as $A^\gamma=A_0^\gamma+ \delta A^\gamma$, where $\gamma=r, a, <$, or $>, A = G$ or $\Sigma$. The frequency-dependent current can thus be expressed as
\begin{widetext}
\begin{equation}
I_\alpha(\omega)=\frac{e}{h}\int_{-\infty}^{+\infty}\frac{\rm dE}{2\pi}Tr[G_0^r(E^+)\delta\Sigma_\alpha^<(E^+,E)
+G_0^<(E^+)\delta\Sigma_\alpha^a(E^+,E)+\delta G^<(E^+,E)\Sigma_{0\alpha}^a(E)
+\delta G^r(E^+,E)\Sigma_{0\alpha}^<(E)+h. c.],
\label{eq:e}
\end{equation}
\end{widetext}
where $E^+=E+\hbar\omega$.
Employing Langreth's rules,\cite{Anantram1995} we obtain
\begin{equation}
\begin{split}
\delta G^<(E_1,E_2)&=G_0^r(E_1)\delta\Sigma_\alpha^<(E_1,E_2)G_0^a(E_2)\\
&+\delta G^r(E_1,E_2)\Sigma_0^<(E_2)G_0^a(E_2)\\
&+G_0^r(E_1)\Sigma_0(E_1)\delta G^a(E_1,E_2).
\label{eq:a}
\end{split}
\end{equation}
And the small-signal terms of the retarded and lesser self-energy are obtained by performing Fourier transform on time-domain expressions\cite{Maciejko2006}
\begin{equation}
\begin{split}
\delta\Sigma_\alpha^<(E^+,E)&=\frac{2\pi iev_\alpha}{2\hbar\omega}[f(E)\Gamma_\alpha(E)-f(E^+)\Gamma_\alpha(E^+)]\\
\delta\Sigma_\alpha^r(E^+,E)&=-\frac{2\pi iev_\alpha}{4\hbar\omega}[\Gamma_\alpha(E)-\Gamma_\alpha(E^+)],
\label{eq:b}
\end{split}
\end{equation}
where $v_\alpha$ is the applied voltage at Lead $\alpha$, $\Gamma_\alpha$ is the coupling between device and Lead $\alpha$, and equilibrium self-energies are $\Sigma_{0\alpha}^r(E)=-i\Gamma_\alpha(E)/2$, $\Sigma_{0\alpha}^<(E)=if(E)\Gamma_\alpha(E)$. The retarded Green's function is expressed by the Dyson equation, $G^r(t,t')=g^r(t,t')+\int{\rm dt_1}{\rm dt_2}G^r(t,t_1)\Sigma^r(t_1,t_2)G^r(t_2,t')$, where $g^r$ is the Green's function for uncoupled the system, and its corresponding small-signal term $\delta g^r$ is the change of $g^r$ due to the change of Hartree potential. Although uncoupled, the device still experiences a potential shift after the application of the external potential; therefore, $\delta g^r$ is nonzero in general, which was neglected in Reference\cite{Anantram1995}. Therefore,
\begin{equation}
\begin{split}
\delta G^r(E_1,E_2)&=G_0^r(E_1)[\left\{{g^r_0(E_1)}\right\}^{-1}\delta g^r(E_1,E_2)\left\{{g_0^r(E_2)}\right\}^{-1}\\
&+\delta\Sigma^r(E_1,E_2)] G_0^r(E_2),
\end{split}
\end{equation}
and $\delta G^a$ is the Hermitian conjugate of $\delta G^r$. $\delta g^r$ can be evaluated from
\begin{equation}
\delta g^r(t_1,t_2)=\int\frac{\rm d\omega}{\omega}U(\omega)(e^{-i\omega t_1}-e^{-i\omega t_2})g_0^r(t_1-t_2),
\end{equation}
where $U(\vec r,\omega)$ is the induced Hartree potential energy in the device region due to the bias voltages, which can be obtained self-consistently by solving Poisson equation subject to the proper boundary condition.\cite{Mo2009} Within the linear-response regime, $U(\vec r, \omega)$ can be approximated as $U(\vec r, \omega)=-\sum_\gamma{eu_\gamma(\vec r, \omega)v_\gamma(\omega)}$, where $u_\gamma(\vec r, t)$ is the dimensionless potential distribution with boundary condition $v_\gamma=1$, and $v_\beta=0$, $(\gamma\ne\beta)$. $\sum_\gamma{u_\gamma(\vec r, \omega)}=1$ is required by gauge invariance. By performing Fourier transform, we have
\begin{equation}
\delta g^r(E^+,E)=\sum_\gamma{\frac{2\pi eu_\gamma(\vec r,\omega)v_\gamma}{2\hbar\omega}[g_0^r(E)-g_0^r(E^+)]}.
\label{eq:c}
\end{equation}
As a consequence, $\delta G^r$ can be expressed as
\begin{equation}
\begin{split}
&\delta G^r(E^+,E)=\sum_\gamma\frac{2\pi ev_\gamma}{2\hbar\omega}G_0^r(E^+)\\
&\times\left[{u_\gamma(\vec r,\omega)\hbar\omega-i\frac{\Gamma_\gamma(E)-\Gamma_\gamma(E^+)}{2}}\right]G_0^r(E).
\label{eq:d}
\end{split}
\end{equation}
Substituting Eqs.~(\ref{eq:a}),~(\ref{eq:b}) and~(\ref{eq:d}) into Eq.~(\ref{eq:e}), we obtain the expression of dynamic admittance for the particle current as:
\begin{widetext}
\begin{equation}
\begin{split}
G_{\alpha\beta}(\omega)&=\frac{ie^2}{h}\int\frac{{\rm d}E}{\hbar\omega}Tr[(f\Gamma_\alpha-\bar f\bar\Gamma_\alpha)(\bar G_0^r-G_0^a)\delta_{\alpha\beta}+\Lambda_{-\alpha}(G_0^<+\bar G_0^<)\delta_{\alpha\beta}+i\Lambda_{+\alpha} \bar G_0^r(f\Gamma_\beta-\bar f\bar\Gamma_\beta)G_0^a\\
&+f\bar G_0^r\left\{{u_\beta(\vec r,\omega)\hbar\omega-i\Lambda_{-\beta}} \right\}G_0^r(\Gamma_\alpha+i\Gamma G_0^a\Lambda_{+\alpha})-\bar f\bar G_0^a\left\{{u_\beta(\vec r,\omega)\hbar\omega+i\Lambda_{-\beta}}\right\}G_0^a(\bar \Gamma_\alpha-i\Lambda_{+\alpha}\bar G_0^r\bar\Gamma)],
\label{eq:main}
\end{split}
\end{equation}
\end{widetext}
where for simplicity, $ A=A(E), \bar A=A(E^+), \Lambda_{-\gamma}=\frac{\Gamma_\gamma(E)-\Gamma_\gamma(E^+)}{2},\Lambda_{+\gamma}=\frac{\Gamma_\gamma(E)+\Gamma_\gamma(E^+)}{2}$, $\Gamma=\sum_\alpha{\Gamma_\alpha}$, and $G_{\alpha\beta}(\omega)=\frac{I_\alpha(\omega)}{v_\beta(\omega)}|_{v_\gamma=0,\gamma\ne\beta}$. Eq.~(\ref{eq:main}) is the central equation of our microscopic theory. The gauge is naturally satisfied as $\sum_\gamma{G_{\alpha\gamma}(\omega)}=0$ based on Eq.~(\ref{eq:main}). In general, current conservation is not satisfied for ac currents through a nano-device; instead, current continuity holds, which can be verified as $\sum_\gamma{G_{\gamma\beta}(\omega)}=i\frac{\omega}{v_\beta}Q(\omega)|_{v_\kappa=0, \beta\ne\kappa}$, and
\begin{equation}
Q(\omega)=\frac{ie}{2\omega}\int_{-\infty}^{+\infty}\frac{\rm dE}{2\pi}Tr[\delta G^<(E^+,E)]
\end{equation}
is the accumulated charge of the device, and is generally nonzero. We have thus developed a gauge-invariant and current-continuous NEGF formalism for ac quantum transport. In practice, we need to solve $u_\gamma(\vec r,\omega)$ self-consistently subjected to the boundary condition $v_\gamma=1$, and $v_\beta=0|_{\beta\ne\gamma}$. When the frequency $\omega$ is low, the WBL approximation yields accurate dynamic admittance. Under the WBL approximation, Eq.~(\ref{eq:main}) can be greatly simplified as
\begin{equation}
\begin{split}
G_{\alpha\beta}(\omega)&=\frac{e^2}{h}\int_{\infty}^{+\infty}\frac{{\rm d}E}{\hbar\omega}Tr[(\Gamma_\alpha\bar G_0^r\Gamma G_0^a-i\hbar\omega\Gamma_\alpha\bar G_0^r G_0^a)\delta_{\alpha\beta}\\
&-\Gamma_\alpha \bar G_0^r\Gamma_\beta G_0^a+i\hbar\omega\bar G_0^ru_\beta(\vec r,\omega)G_0^a\Gamma_\alpha](f-\bar f).
\label{eq:wbl}
\end{split}
\end{equation}
Eq.~(\ref{eq:wbl}) is consistent with the formula for the dynamic admittance of the total current in Reference\cite{Wei2009}. The last term in Eq.~(\ref{eq:wbl}) that is proportional to $u_\beta(\vec r,\omega)$ was mistakenly termed as displacement current, and the rest as the particle current. In fact, all terms are for the particle current.

According to Reference\cite{Wang1999}, within the WBL approximation, the dynamic admittance can be
expressed as
\begin{equation}
\begin{split}
G_{\alpha\beta}^{W}(\omega)&=\frac{e^2}{h}\int_{-\infty}^{+\infty}\frac{{\rm d}E}{\hbar\omega}Tr[(\Gamma_\alpha\bar G_0^r\Gamma G_0^a-i\hbar\omega\Gamma_\alpha\bar G_0^r G_0^a)\delta_{\alpha\beta}\\
&-\Gamma_\alpha \bar G_0^r\Gamma_\beta G_0^a+iA_\alpha\hbar\omega\bar G_0^rG_0^a\Gamma_\beta](f-\bar f),
\label{eq:wang}
\end{split}
\end{equation}where the first three terms are the results of Anantram and Datta,\cite{Anantram1995} and the fourth is the phenomenological term that was thought to count for the contribution of the displacement current. Comparing Eqs.~(\ref{eq:wbl}) and~(\ref{eq:wang}), we find that the two are exactly the same when
\begin{equation}
u_\beta(\vec r,\omega)=A_\beta,
\label{eq:u}
\end{equation}
where $A_\beta=\frac{\int_{-\infty}^{+\infty}{\rm dE}Tr(\bar G_0^r\Gamma_\beta G_0^a)(f-\bar f)}{\int_{-\infty}^{+\infty}{\rm dE}Tr(\bar G_0^r\Gamma G_0^a)(f-\bar f)}$, namely, the device is a perfect conductor with a uniform potential at
\begin{equation}
V=\sum_\gamma{A_\gamma v_\gamma}.
\label{eq:v}
\end{equation}
In other words, Eq.~(\ref{eq:wang}) is recovered only if the device is a perfect conductor with its uniform potential expressed by Eq.~(\ref{eq:v}). As Eq.~(\ref{eq:wbl}) is for the particle current, the fourth term in Eq.~(\ref{eq:wang}) is thus not from the displacement current, and is rather the correction to the particle current due to the induced potential of the device. It is important to emphasize this.

Employing Eq.~(\ref{eq:wbl}), we calculate the dynamic admittance of a (5, 5) carbon nanotube coupled with two aluminum electrodes as depicted in Fig. 1a. The result is compared to that of TDDFT-NEGF calculation.\cite{Zheng2007,Zheng2010} TDDFT-NEGF has been developed to calculate the time-dependent current, and employed to simulate the transient current through a variety of molecular and nanoscopic devices such as carbon nanotube based two-terminal device.\cite{Yam2008} The local density approximation (LDA) and WBL approximation are adopted. The time-dependent current is evaluated, and a Fourier transform is performed to determine the dynamic admittance. The real and imaginary parts of $G_{LL}(\omega)$ are plotted in Fig.1b. Clearly, there is an excellent agreement between the frequency- and time-domain results, which confirms the validity of our gauge-invariant and current-continuous ac quantum transport theory.
\begin{figure}
\includegraphics[width=7.5cm]{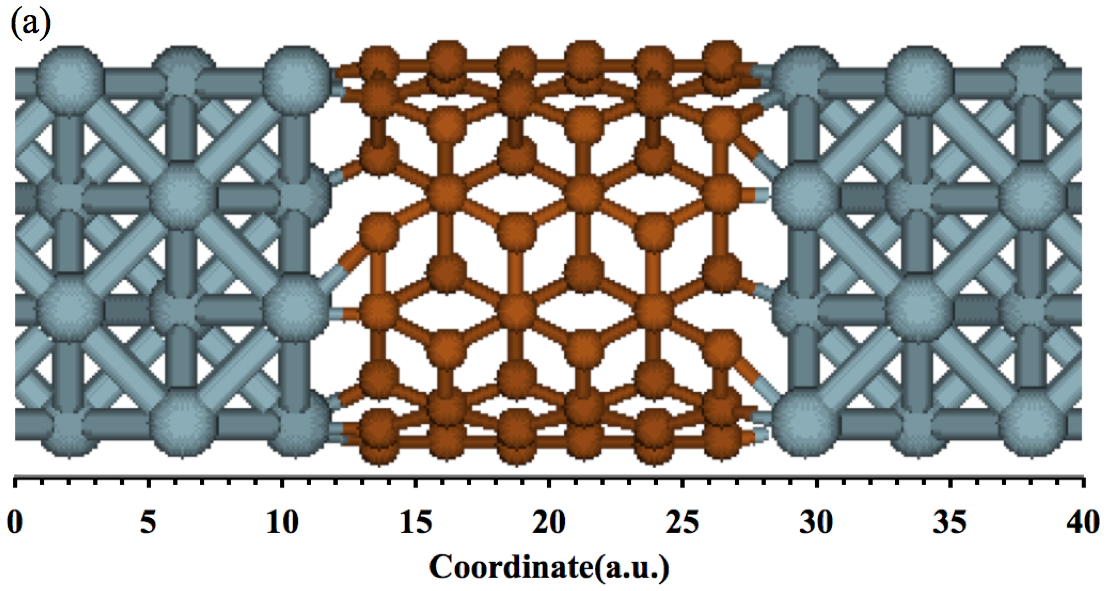}
\includegraphics[width=7.5cm]{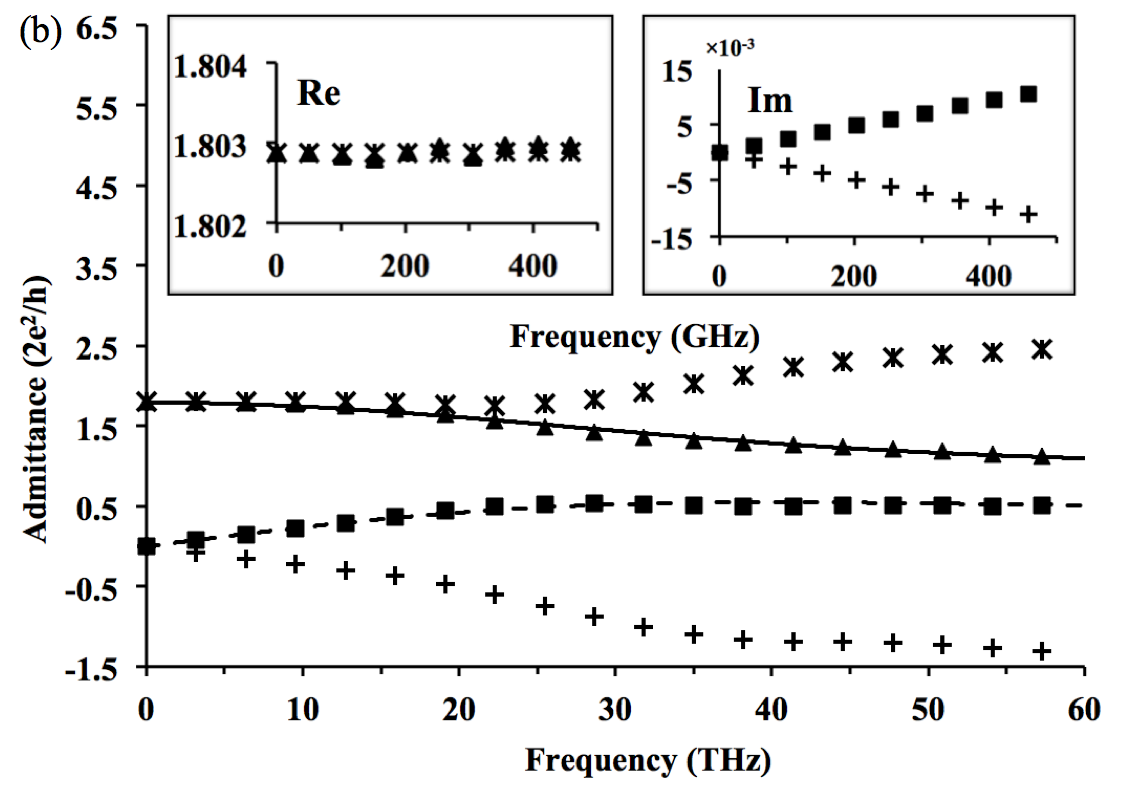}
\caption{\label{fig.1}(a) The structure of a (5, 5) CNT embedded with aluminum electrodes. The coordinate indicates the positions of atoms in real space. (b) Results from the three different methods for an inhomogeneous Al-CNT-Al system. Solid (dashed) line: real (imaginary) part of the dynamic admittance from TDDFT-NEGF calculation; triangles (squares): real (imaginary) part from Eq.~(\ref{eq:wbl}); asterisks (crosses): real (imaginary) part from the phenomenological formula in Reference\cite{Wang1999}. Two smaller figures are the real (Re) and imaginary (Im) parts of both ac transport theories below 500 GHz.}
\end{figure}

In comparison, we calculate the corresponding dynamic admittances of both devices employing the phenomenological formula given in Reference\cite{Wang1999}, and the resulting $G_{LL}(\omega)$ is plotted in Fig.1b, as well. When the frequency $\omega$ is zero, the phenomenological method recovers both the results of the  microscopic theory and TDDFT-NEGF. However, beyond the steady state, the results deviate from those of Eq.~(\ref{eq:wbl}) and the TDDFT-NEGF.\cite{Zheng2007,Zheng2010} In particular, as in Fig.1b, the imaginary part of dynamic admittance starts to deviate immediately at $\omega>0$. At both low and high frequencies, the potential drops mostly across the CNT, which differs drastically from the constant potential requirement of Eqs.~(\ref{eq:u}) and (\ref{eq:v}). In fact, if the device and electrodes are mirror-symmetric, then $A_L=A_R=\frac{1}{2}$, and $V=\frac{1}{2}V_L$ (with $V_R=0$). Certain constant potential distribution $u_L(\vec r,\omega)=\frac{1}{2}$ would have led to the agreement of Eqs.~(\ref{eq:wbl}) and (\ref{eq:wang}). However, this is not the case.
\begin{figure}
\includegraphics[width=7.5cm]{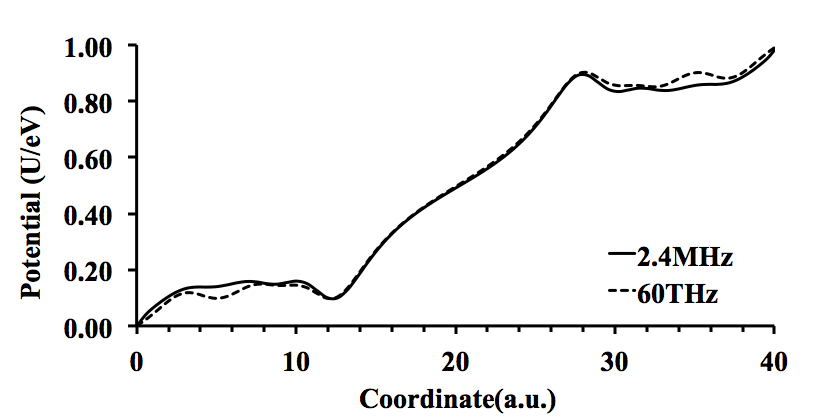}
\caption{\label{fig.2}Potential distribution [$u_R(\vec r,\omega)$] in the device region of the Al-CNT-Al system.}
\end{figure}

We explore how to include the contribution of the displacement current. Following the treatment of B\"{u}ttiker and coworkers,\cite{Buttiker1993} the admittance for the displacement current can be defined as
\begin{equation}
G_\beta^d(\omega)=\frac{I^d(\omega)}{v_\beta}=\frac{-\sum_\alpha{I_\alpha(\omega)}}{v_\beta}=-\sum_\alpha{G_{\alpha\beta}(\omega)},
\end{equation}
where $v_\kappa=0, \kappa\ne\beta$. $\sum_\beta{G_\beta^d}=-\sum_\beta\sum_\alpha{G_{\alpha\beta}}=0$ ensures that the gauge invariance is satisfied. We may define the total admittance as $G_{\alpha\beta}^{tot}=G_{\alpha\beta}+A_\alpha G_\beta^d$. Gauge invariance and current conservation for $G^{tot}$ are satisfied as long as $\sum_\alpha A_\alpha=1$. This means that there are infinite ways to partition $G_\beta^d$ if \textsl{one is merely to ensure the gauge invariance and current conservation for the dynamic admittance of the total current}. Although the partition of $G^d$ in Reference\cite{Wang1999} was uniquely determined, it was based on the expression that violates gauge invariance for the dynamic admittance of the particle current. If the correct expression, Eq.~(\ref{eq:main}), is used, the correction to the dynamic admittance would simply be zero! According to classical electrodynamics, the displacement current is introduced to ensure the current conservation, and is physically defined. According to Amp\`{e}re's law, the displacement current should be $I^d(\omega)=i\omega\Phi(\omega)$ , where $\Phi(\omega)$ is electric flux. The total current is thus $I_\alpha^{tot}=I_\alpha+i\omega\Phi_\alpha$, where $\Phi_\alpha$ is the flux of the interface between Lead $\alpha$ and device, i.e. $A_\alpha=\Phi_\alpha/\Phi$. We have thus an expression for the admittance from the displacement current. However, the calculation of the contribution from the displacement current can be numerically difficult, because normally the portions of the electrodes are included in the simulation box. As a result, the electric fields at the boundaries of the electrodes are very small, and $\Phi_\alpha$ and $\Phi$ may thus be very close to zero. The evaluation of $A_\alpha$ may encounter the $0^{\pm}/0^{\pm}$ problem, which renders the accurate evaluation of $A_\alpha$ difficult. Moreover, $A_\alpha$ depends sensitively on the exact location of the boundary. Alternatively, inspired by B\"{u}ttiker \textsl{et al} in Reference\cite{Buttiker1993}, we introduce Terminal 0, with its dynamic admittance component expressed as $G_{0\beta}(\omega)=G_\beta^d(\omega)=-\sum_\alpha{G_{\alpha\beta}(\omega)}$. As a result, the total curent conserves, since $\sum_{\alpha=0}{G_{\alpha\beta}}=0$

The microscopic ac quantum transport theory developed here is gauge-invariant but not current-conserved. $Q(\omega)$ ￼ depends on the size of the device region. As more of the leads are included in the simulation box, $Q(\omega)$ becomes less. In fact, if the simulation box is infinitely long and includes infinite portion of the leads, $Q(\omega)$ ￼ is zero; thus, current conservation holds. However, since the simulation box is finite in practice, $Q(\omega)$ is usually nonzero, and current conservation is thus not guaranteed.
\begin{acknowledgments}
We would like to thank Prof. Jian Wang and Prof. Hong Guo for helpful discussion. The financial support from the Hong Kong Research Grant Council (HKU 700808P, HKU 700909P, HKU 700711P, HKUST9/CRF/08, AoE/P-04/08), the Natural Science Foundation of China (Nos. 21103157 and 21233007), and the Fundamental Research Funds for Central Universities (Nos. 2340000025 and 2340000034) is gratefully acknowledged.
\end{acknowledgments}

\bibliography{GICC}

\begin{thebibliography}{18}%
\makeatletter
\providecommand \@ifxundefined [1]{%
 \@ifx{#1\undefined}
}%
\providecommand \@ifnum [1]{%
 \ifnum #1\expandafter \@firstoftwo
 \else \expandafter \@secondoftwo
 \fi
}%
\providecommand \@ifx [1]{%
 \ifx #1\expandafter \@firstoftwo
 \else \expandafter \@secondoftwo
 \fi
}%
\providecommand \natexlab [1]{#1}%
\providecommand \enquote  [1]{``#1''}%
\providecommand \bibnamefont  [1]{#1}%
\providecommand \bibfnamefont [1]{#1}%
\providecommand \citenamefont [1]{#1}%
\providecommand \href@noop [0]{\@secondoftwo}%
\providecommand \href [0]{\begingroup \@sanitize@url \@href}%
\providecommand \@href[1]{\@@startlink{#1}\@@href}%
\providecommand \@@href[1]{\endgroup#1\@@endlink}%
\providecommand \@sanitize@url [0]{\catcode `\\12\catcode `\$12\catcode
  `\&12\catcode `\#12\catcode `\^12\catcode `\_12\catcode `\%12\relax}%
\providecommand \@@startlink[1]{}%
\providecommand \@@endlink[0]{}%
\providecommand \url  [0]{\begingroup\@sanitize@url \@url }%
\providecommand \@url [1]{\endgroup\@href {#1}{\urlprefix }}%
\providecommand \urlprefix  [0]{URL }%
\providecommand \Eprint [0]{\href }%
\providecommand \doibase [0]{http://dx.doi.org/}%
\providecommand \selectlanguage [0]{\@gobble}%
\providecommand \bibinfo  [0]{\@secondoftwo}%
\providecommand \bibfield  [0]{\@secondoftwo}%
\providecommand \translation [1]{[#1]}%
\providecommand \BibitemOpen [0]{}%
\providecommand \bibitemStop [0]{}%
\providecommand \bibitemNoStop [0]{.\EOS\space}%
\providecommand \EOS [0]{\spacefactor3000\relax}%
\providecommand \BibitemShut  [1]{\csname bibitem#1\endcsname}%
\let\auto@bib@innerbib\@empty
\bibitem [{\citenamefont {Xue}\ \emph {et~al.}(2002)\citenamefont {Xue},
  \citenamefont {Datta},\ and\ \citenamefont {Ratner}}]{Xue2002}%
  \BibitemOpen
  \bibfield  {author} {\bibinfo {author} {\bibfnamefont {Y.~Q.}\ \bibnamefont
  {Xue}}, \bibinfo {author} {\bibfnamefont {S.}~\bibnamefont {Datta}}, \ and\
  \bibinfo {author} {\bibfnamefont {M.~A.}\ \bibnamefont {Ratner}},\ }\href
  {\doibase 10.1016/S0301-0104(02)00446-9} {\bibfield  {journal} {\bibinfo
  {journal} {Chem. Phys.}\ }\textbf {\bibinfo {volume} {281}},\ \bibinfo
  {pages} {151} (\bibinfo {year} {2002})}\BibitemShut {NoStop}%
\bibitem [{\citenamefont {Datta}(2004)}]{Datta2004}%
  \BibitemOpen
  \bibfield  {author} {\bibinfo {author} {\bibfnamefont {S.}~\bibnamefont
  {Datta}},\ }\href {\doibase 10.1088/0957-4484/15/7/051} {\bibfield  {journal}
  {\bibinfo  {journal} {Nanotechnology}\ }\textbf {\bibinfo {volume} {15}},\
  \bibinfo {pages} {S433} (\bibinfo {year} {2004})}\BibitemShut {NoStop}%
\bibitem [{\citenamefont {Taylor}\ \emph {et~al.}(2001)\citenamefont {Taylor},
  \citenamefont {Guo},\ and\ \citenamefont {Wang}}]{Taylor2001}%
  \BibitemOpen
  \bibfield  {author} {\bibinfo {author} {\bibfnamefont {J.}~\bibnamefont
  {Taylor}}, \bibinfo {author} {\bibfnamefont {H.}~\bibnamefont {Guo}}, \ and\
  \bibinfo {author} {\bibfnamefont {J.}~\bibnamefont {Wang}},\ }\href {\doibase
  10.1103/PhysRevB.63.245407} {\bibfield  {journal} {\bibinfo  {journal} {Phys.
  Rev. B}\ }\textbf {\bibinfo {volume} {63}},\ \bibinfo {pages} {245407}
  (\bibinfo {year} {2001})}\BibitemShut {NoStop}%
\bibitem [{\citenamefont {Lang}(1995)}]{Lang1995}%
  \BibitemOpen
  \bibfield  {author} {\bibinfo {author} {\bibfnamefont {N.~D.}\ \bibnamefont
  {Lang}},\ }\href {\doibase 10.1103/PhysRevB.52.5335} {\bibfield  {journal}
  {\bibinfo  {journal} {Phys. Rev. B}\ }\textbf {\bibinfo {volume} {52}},\
  \bibinfo {pages} {5335} (\bibinfo {year} {1995})}\BibitemShut {NoStop}%
\bibitem [{\citenamefont {Lang}\ and\ \citenamefont
  {Avouris}(2000)}]{Lang2000}%
  \BibitemOpen
  \bibfield  {author} {\bibinfo {author} {\bibfnamefont {N.~D.}\ \bibnamefont
  {Lang}}\ and\ \bibinfo {author} {\bibfnamefont {P.}~\bibnamefont {Avouris}},\
  }\href {\doibase 10.1103/PhysRevLett.84.358} {\bibfield  {journal} {\bibinfo
  {journal} {Phys. Rev. Lett.}\ }\textbf {\bibinfo {volume} {84}},\ \bibinfo
  {pages} {358} (\bibinfo {year} {2000})}\BibitemShut {NoStop}%
\bibitem [{\citenamefont {Ke}\ \emph {et~al.}(2004)\citenamefont {Ke},
  \citenamefont {Baranger},\ and\ \citenamefont {Yang}}]{Ke2004}%
  \BibitemOpen
  \bibfield  {author} {\bibinfo {author} {\bibfnamefont {S.~H.}\ \bibnamefont
  {Ke}}, \bibinfo {author} {\bibfnamefont {H.~U.}\ \bibnamefont {Baranger}}, \
  and\ \bibinfo {author} {\bibfnamefont {W.~T.}\ \bibnamefont {Yang}},\ }\href
  {\doibase 10.1021/ja047367e} {\bibfield  {journal} {\bibinfo  {journal} {J.
  Am. Chem. Soc.}\ }\textbf {\bibinfo {volume} {126}},\ \bibinfo {pages}
  {15897} (\bibinfo {year} {2004})}\BibitemShut {NoStop}%
\bibitem [{\citenamefont {B{\"u}ttiker}\ \emph {et~al.}(1993)\citenamefont
  {B{\"u}ttiker}, \citenamefont {Pr{\^e}tre},\ and\ \citenamefont
  {Thomas}}]{Buttiker1993}%
  \BibitemOpen
  \bibfield  {author} {\bibinfo {author} {\bibfnamefont {M.}~\bibnamefont
  {B{\"u}ttiker}}, \bibinfo {author} {\bibfnamefont {A.}~\bibnamefont
  {Pr{\^e}tre}}, \ and\ \bibinfo {author} {\bibfnamefont {H.}~\bibnamefont
  {Thomas}},\ }\href {\doibase 10.1103/PhysRevLett.70.4114} {\bibfield
  {journal} {\bibinfo  {journal} {Phys. Rev. Lett.}\ }\textbf {\bibinfo
  {volume} {70}},\ \bibinfo {pages} {4114} (\bibinfo {year}
  {1993})}\BibitemShut {NoStop}%
\bibitem [{\citenamefont {Anantram}\ and\ \citenamefont
  {Datta}(1995)}]{Anantram1995}%
  \BibitemOpen
  \bibfield  {author} {\bibinfo {author} {\bibfnamefont {M.~P.}\ \bibnamefont
  {Anantram}}\ and\ \bibinfo {author} {\bibfnamefont {S.}~\bibnamefont
  {Datta}},\ }\href {\doibase 10.1103/PhysRevB.51.7632} {\bibfield  {journal}
  {\bibinfo  {journal} {Phys. Rev. B}\ }\textbf {\bibinfo {volume} {51}},\
  \bibinfo {pages} {7632} (\bibinfo {year} {1995})}\BibitemShut {NoStop}%
\bibitem [{\citenamefont {Wang}\ \emph {et~al.}(1999)\citenamefont {Wang},
  \citenamefont {Wang},\ and\ \citenamefont {Guo}}]{Wang1999}%
  \BibitemOpen
  \bibfield  {author} {\bibinfo {author} {\bibfnamefont {B.~G.}\ \bibnamefont
  {Wang}}, \bibinfo {author} {\bibfnamefont {J.}~\bibnamefont {Wang}}, \ and\
  \bibinfo {author} {\bibfnamefont {H.}~\bibnamefont {Guo}},\ }\href {\doibase
  10.1103/PhysRevLett.82.398} {\bibfield  {journal} {\bibinfo  {journal} {Phys.
  Rev. Lett.}\ }\textbf {\bibinfo {volume} {82}},\ \bibinfo {pages} {398}
  (\bibinfo {year} {1999})}\BibitemShut {NoStop}%
\bibitem [{\citenamefont {Yamamoto}\ \emph {et~al.}(2010)\citenamefont
  {Yamamoto}, \citenamefont {Sasaoka}, \citenamefont {Watanabe},\ and\
  \citenamefont {Watanabe}}]{Yamamoto2010}%
  \BibitemOpen
  \bibfield  {author} {\bibinfo {author} {\bibfnamefont {T.}~\bibnamefont
  {Yamamoto}}, \bibinfo {author} {\bibfnamefont {K.}~\bibnamefont {Sasaoka}},
  \bibinfo {author} {\bibfnamefont {S.}~\bibnamefont {Watanabe}}, \ and\
  \bibinfo {author} {\bibfnamefont {K.}~\bibnamefont {Watanabe}},\ }\href
  {\doibase 10.1103/PhysRevB.81.115448} {\bibfield  {journal} {\bibinfo
  {journal} {Phys. Rev. B}\ }\textbf {\bibinfo {volume} {81}},\ \bibinfo
  {pages} {115448} (\bibinfo {year} {2010})}\BibitemShut {NoStop}%
\bibitem [{\citenamefont {Zheng}\ \emph {et~al.}(2007)\citenamefont {Zheng},
  \citenamefont {Wang}, \citenamefont {Yam}, \citenamefont {Mo},\ and\
  \citenamefont {Chen}}]{Zheng2007}%
  \BibitemOpen
  \bibfield  {author} {\bibinfo {author} {\bibfnamefont {X.}~\bibnamefont
  {Zheng}}, \bibinfo {author} {\bibfnamefont {F.}~\bibnamefont {Wang}},
  \bibinfo {author} {\bibfnamefont {C.~Y.}\ \bibnamefont {Yam}}, \bibinfo
  {author} {\bibfnamefont {Y.}~\bibnamefont {Mo}}, \ and\ \bibinfo {author}
  {\bibfnamefont {G.~H.}\ \bibnamefont {Chen}},\ }\href {\doibase
  10.1103/PhysRevB.75.195127} {\bibfield  {journal} {\bibinfo  {journal} {Phys.
  Rev. B}\ }\textbf {\bibinfo {volume} {75}},\ \bibinfo {pages} {195127}
  (\bibinfo {year} {2007})}\BibitemShut {NoStop}%
\bibitem [{\citenamefont {Zheng}\ \emph {et~al.}(2010)\citenamefont {Zheng},
  \citenamefont {Chen}, \citenamefont {Mo}, \citenamefont {Koo}, \citenamefont
  {Tian}, \citenamefont {Yam},\ and\ \citenamefont {Yan}}]{Zheng2010}%
  \BibitemOpen
  \bibfield  {author} {\bibinfo {author} {\bibfnamefont {X.}~\bibnamefont
  {Zheng}}, \bibinfo {author} {\bibfnamefont {G.~H.}\ \bibnamefont {Chen}},
  \bibinfo {author} {\bibfnamefont {Y.}~\bibnamefont {Mo}}, \bibinfo {author}
  {\bibfnamefont {S.~K.}\ \bibnamefont {Koo}}, \bibinfo {author} {\bibfnamefont
  {H.}~\bibnamefont {Tian}}, \bibinfo {author} {\bibfnamefont {C.~Y.}\
  \bibnamefont {Yam}}, \ and\ \bibinfo {author} {\bibfnamefont {Y.~J.}\
  \bibnamefont {Yan}},\ }\href {\doibase 10.1063/1.3475566} {\bibfield
  {journal} {\bibinfo  {journal} {J. Chem. Phys.}\ }\textbf {\bibinfo {volume}
  {133}},\ \bibinfo {pages} {114101} (\bibinfo {year} {2010})}\BibitemShut
  {NoStop}%
\bibitem [{\citenamefont {Zhuang}\ \emph {et~al.}(2011)\citenamefont {Zhuang},
  \citenamefont {Zhang},\ and\ \citenamefont {Wang}}]{Zhuang2011}%
  \BibitemOpen
  \bibfield  {author} {\bibinfo {author} {\bibfnamefont {J.~N.}\ \bibnamefont
  {Zhuang}}, \bibinfo {author} {\bibfnamefont {L.}~\bibnamefont {Zhang}}, \
  and\ \bibinfo {author} {\bibfnamefont {J.}~\bibnamefont {Wang}},\ }\href
  {\doibase 10.1063/1.3673566} {\bibfield  {journal} {\bibinfo  {journal} {AIP
  Advances}\ }\textbf {\bibinfo {volume} {1}},\ \bibinfo {pages} {042180}
  (\bibinfo {year} {2011})}\BibitemShut {NoStop}%
\bibitem [{\citenamefont {Wei}\ and\ \citenamefont {Wang}(2009)}]{Wei2009}%
  \BibitemOpen
  \bibfield  {author} {\bibinfo {author} {\bibfnamefont {Y.~D.}\ \bibnamefont
  {Wei}}\ and\ \bibinfo {author} {\bibfnamefont {J.}~\bibnamefont {Wang}},\
  }\href {\doibase 10.1103/PhysRevB.79.195315} {\bibfield  {journal} {\bibinfo
  {journal} {Phys. Rev. B}\ }\textbf {\bibinfo {volume} {79}},\ \bibinfo
  {pages} {195315} (\bibinfo {year} {2009})}\BibitemShut {NoStop}%
\bibitem [{\citenamefont {Jauho}\ \emph {et~al.}(1994)\citenamefont {Jauho},
  \citenamefont {Wingreen},\ and\ \citenamefont {Meir}}]{Jauho1994}%
  \BibitemOpen
  \bibfield  {author} {\bibinfo {author} {\bibfnamefont {A.~P.}\ \bibnamefont
  {Jauho}}, \bibinfo {author} {\bibfnamefont {N.~S.}\ \bibnamefont {Wingreen}},
  \ and\ \bibinfo {author} {\bibfnamefont {Y.}~\bibnamefont {Meir}},\ }\href
  {\doibase 10.1103/PhysRevB.50.5528} {\bibfield  {journal} {\bibinfo
  {journal} {Phys. Rev. B}\ }\textbf {\bibinfo {volume} {50}},\ \bibinfo
  {pages} {5528} (\bibinfo {year} {1994})}\BibitemShut {NoStop}%
\bibitem [{\citenamefont {Maciejko}\ \emph {et~al.}(2006)\citenamefont
  {Maciejko}, \citenamefont {Wang},\ and\ \citenamefont {Guo}}]{Maciejko2006}%
  \BibitemOpen
  \bibfield  {author} {\bibinfo {author} {\bibfnamefont {J.}~\bibnamefont
  {Maciejko}}, \bibinfo {author} {\bibfnamefont {J.}~\bibnamefont {Wang}}, \
  and\ \bibinfo {author} {\bibfnamefont {H.}~\bibnamefont {Guo}},\ }\href
  {\doibase 10.1103/PhysRevB.74.085324} {\bibfield  {journal} {\bibinfo
  {journal} {Phys. Rev. B}\ }\textbf {\bibinfo {volume} {74}},\ \bibinfo
  {pages} {085324} (\bibinfo {year} {2006})}\BibitemShut {NoStop}%
\bibitem [{\citenamefont {Mo}\ \emph {et~al.}(2009)\citenamefont {Mo},
  \citenamefont {Zheng}, \citenamefont {Chen},\ and\ \citenamefont
  {Yan}}]{Mo2009}%
  \BibitemOpen
  \bibfield  {author} {\bibinfo {author} {\bibfnamefont {Y.}~\bibnamefont
  {Mo}}, \bibinfo {author} {\bibfnamefont {X.}~\bibnamefont {Zheng}}, \bibinfo
  {author} {\bibfnamefont {G.~H.}\ \bibnamefont {Chen}}, \ and\ \bibinfo
  {author} {\bibfnamefont {Y.~J.}\ \bibnamefont {Yan}},\ }\href {\doibase
  10.1088/0953-8984/21/35/355301} {\bibfield  {journal} {\bibinfo  {journal}
  {J. Phys.: Condens. Matter}\ }\textbf {\bibinfo {volume} {21}},\ \bibinfo
  {pages} {355301} (\bibinfo {year} {2009})}\BibitemShut {NoStop}%
\bibitem [{\citenamefont {Yam}\ \emph {et~al.}(2008)\citenamefont {Yam},
  \citenamefont {Mo}, \citenamefont {Wang}, \citenamefont {Li}, \citenamefont
  {Chen}, \citenamefont {Zheng}, \citenamefont {Matsuda}, \citenamefont
  {Tahir-Kheli},\ and\ \citenamefont {Goddard}}]{Yam2008}%
  \BibitemOpen
  \bibfield  {author} {\bibinfo {author} {\bibfnamefont {C.~Y.}\ \bibnamefont
  {Yam}}, \bibinfo {author} {\bibfnamefont {Y.}~\bibnamefont {Mo}}, \bibinfo
  {author} {\bibfnamefont {F.}~\bibnamefont {Wang}}, \bibinfo {author}
  {\bibfnamefont {X.~B.}\ \bibnamefont {Li}}, \bibinfo {author} {\bibfnamefont
  {G.~H.}\ \bibnamefont {Chen}}, \bibinfo {author} {\bibfnamefont
  {X.}~\bibnamefont {Zheng}}, \bibinfo {author} {\bibfnamefont
  {Y.}~\bibnamefont {Matsuda}}, \bibinfo {author} {\bibfnamefont
  {J.}~\bibnamefont {Tahir-Kheli}}, \ and\ \bibinfo {author} {\bibfnamefont
  {W.~A.}\ \bibnamefont {Goddard}},\ }\href {\doibase
  10.1088/0957-4484/19/49/495203} {\bibfield  {journal} {\bibinfo  {journal}
  {Nanotechnology}\ }\textbf {\bibinfo {volume} {19}},\ \bibinfo {pages}
  {495203} (\bibinfo {year} {2008})}\BibitemShut {NoStop}%
\end{thebibliography}%
$^*$Email: ghc@everest.hku.hk
\end{document}